\documentclass[referee,A4]{aa} 
\usepackage{psfig,amssymb,alltt,epsfig}
\usepackage{latexsym}
\usepackage{graphicx}
\graphicspath{{PS/}}
\usepackage{psfig}
\psfigurepath{./PS:.}
\usepackage[english]{babel}
\usepackage{indentfirst}
\usepackage{amsmath}
\usepackage{natbib}

\psfigurepath{./PS}
\vspace{1cm}
\newcommand{\be}{\begin{eqnarray}}
\newcommand{\ee}{\end{eqnarray}}

\newcommand{\bitem}{\begin{itemize}}
\newcommand{\eitem}{\end{itemize}}

\def\inv{^{-1}}
\def\adj{^\dagger }

\begin{document}

\title{Wavelets, ridgelets and curvelets on the sphere}

\author{Jean-Luc Starck \inst{1}\inst{2} \and Yassir Moudden \inst{1} \and Pierrick Abrial \inst{1} \and Mai Nguyen \inst{3} }

\institute{ 
\inst{1} DAPNIA/SEDI-SAP, Service d'Astrophysique, CEA-Saclay, 
 F-91191 Gif-sur-Yvette Cedex, France. \\
 \inst{2} Department of Statistics, Sequoia Hall, 390 Serra Mall, \\
 Stanford University, Stanford, CA 94305 USA. \\
\inst{3} Laboratoire Traitement de l'Image et du Signal, 
   CNRS UMR 8051/ENSEA/Universit\'e de Cergy-Pontoise, 
    ENSEA 6 avenue du ponceau, 95014 Cergy.}

\offprints{jstarck@cea.fr}
 
\date{\today}

 

\abstract{We present in this paper new multiscale transforms on the sphere, namely the isotropic
undecimated wavelet transform, the pyramidal wavelet transform,
the ridgelet transform and the curvelet transform. All of these transforms can be inverted 
i.e. we can exactly reconstruct the original data from its coefficients in either representation. 
Several applications are described. 
We show how these transforms can be used in denoising and especially in a  Combined Filtering Method, 
which uses both the wavelet and the curvelet transforms, 
thus benefiting from the advantages of both transforms. 
An application to component separation from multichannel data mapped to the sphere is also 
described in which we take advantage of moving to a wavelet representation.}

\maketitle 
\markboth{Wavelets, Ridgelets and Curvelets on the Sphere}{}


\section{Introduction}
\subsection*{Wavelets in Astronomy}
Wavelets are very popular tools in astronomy \citep{starck:book02} which have led to very impressive
results in denoising and detection applications. For instance, both the Chandra and the XMM data centers use wavelets
for the detection of extended sources in X-ray images. For denoising and deconvolution, wavelets have also 
demonstrated how powerful they are for discriminating signal from noise \citep{starck:sta02_2}.
{ In cosmology, wavelets have been used in many studies such as 
for analyzing the spatial 
distribution of galaxies \citep{astro:slezak93,astro:escalera95,starck:sta05,starck:martinez05}, 
determining the topology of the universe \citep{astro:rocha04}, detecting non-Gaussianity
in the CMB maps \citep{gauss:aghanim99,gauss:barreiro01_1,wave:vielva04,starck:sta03_1},
reconstructing the primordial power spectrum \citep{astro:pia03},
measuring the galaxy power spectrum  \citep{astro:fang00} or reconstructing weak lensing
mass maps \citep{starck:sta05b}.
It has also been shown that noise is a problem of major concern for N-body simulations 
of structure formation in the 
early Universe and that using wavelets
to remove noise from N-body simulations is equivalent to 
simulations with two orders 
of magnitude more particles \citep{rest:romeo03,rest:romeo04}. 
}

The most popular 
wavelet algorithm in astrophysical applications 
is the so-called ``\emph{\`a trous} algorithm''. It is an  isotropic undecimated
wavelet transform in which the scaling function used is a Box-Spline of order three.
The isotropy of the wavelet function makes this decomposition optimal for the detection
of isotropic objects. The non decimation makes the decomposition redundant (the number of coefficients
in the decomposition is equal to the number of samples in the data multiplied by the number of scales)
and allows us to avoid Gibbs aliasing after reconstruction in image restoration applications, 
as generally occurs  with orthogonal or bi-orthogonal wavelet transforms. The choice of a $B_3$-spline is motivated by the fact that we 
want an analyzing function close to a Gaussian, but verifying the dilation equation, which is 
required in order to have a fast transformation. One last property of this algorithm is to provide
a very straightforward reconstruction. Indeed, the sum of all the wavelet scales and of the coarsest resolution
image reproduces exactly the original image. 

When analyzing data that contains anisotropic features, wavelets are no longer optimal and this
has motivated the development of new multiscale decompositions such as the ridgelet and the curvelet transforms
\citep{cur:donoho99,starck:sta01_3}. In Starck et al. \citep*{starck:sta03_1}, it has been shown that
the curvelet transform could be useful for the detection and the discrimination 
of non Gaussianity in CMB data. In this area, further insight will come from the analysis of full-sky data mapped to the sphere thus  requiring the development of a curvelet transform on the sphere. 
 
\subsection*{Wavelets on the sphere}
Several wavelet transforms on the sphere have been proposed in recent years. 
 Schr\"oder and  Sweldens \citep*{wave:sweldens95a} have developed an orthogonal wavelet 
 transform on the sphere based on the Haar wavelet function.
 Its interest is however relatively limited because of the poor properties of the Haar function and the
 problems inherent to the orthogonal decomposition. Many papers describe 
 new continuous wavelet transforms \citep{wave:antoine99,wave:tenerio99,wave:cayon01,wave:holschneider96} and the recent 
 detection of non-Gaussianity in CMB was obtained by Vielva et al. \citep{wave:vielva04} using the
 continuous Mexican Hat wavelet transform \citep{wave:cayon01}. These works have been extended to directional
 wavelet transforms \citep{wave:antoine01,wave:hobson04,wave:wiaux}. All these new continuous wavelet decompositions 
 are interesting for data analysis, but cannot be used for restoration purposes because of the lack 
 of an inverse transform. Only the algorithm proposed by Freeden and Maier \citep{freeden97,freeden98}, 
 based on the Spherical Harmonic Decomposition, has an inverse transform.  


The goal of this paper is to extend existing 2D multiscale decompositions, namely the ridgelet 
transform, the curvelet transform and the isotropic wavelet transform,  which work on flat images to 
multiscale methods on the sphere. In section~\ref{sect_wts}, we present a new isotropic wavelet
transform on the sphere which has similar properties to the \emph{\`a trous} algorithm and therefore should
be very useful for data denoising and deconvolution. This algorithm is directly derived
from the FFT-based wavelet transform proposed in \citet{starck:sta94_3} for aperture synthesis
image restoration, and is relatively close to the Freeden and Maier \citep{freeden98} method,
except that it features the same straightforward reconstruction as the \emph{\`a trous} algorithm 
(i.e. the sum of the scales reproduces the original data). This new wavelet transform also
can be easily  extended  to a pyramidal wavelet transform, hence with reduced redundancy, a possibility
which may be very important for larger data sets such as expected from the future Planck-Surveyor experiment.   
In section~\ref{sect_cur}, we show how this new decomposition can be used to derive a curvelet transform on the sphere.
Section~\ref{sect_exp} describes how these new transforms can be used in denoising applications and introduces
the Combined Filtering Method, which allows us to filter data on the sphere 
using both the Wavelet and the Curvelet transforms.
In section~\ref{sect_ica}, we show that the independent component analysis method wSMICA \citep{starck:yassir05},
which was designed for 2D data, can be extended to data on the sphere.
 
\section{Wavelet transform on the sphere}
\label{sect_wts}
\subsection{Isotropic Undecimated Wavelet Transform on the Sphere (UWTS) }
There are clearly many different possible implementations of a wavelet transform 
on the sphere and their performance depends on the application. 
We describe here an undecimated isotropic transform which has many properties in common with
the \emph{ \`a trous} algorithm, and is therefore a good candidate for restoration applications.
Its isotropy is a favorable property when analyzing  a 
statistically isotropic Gaussian field such as the CMB or data sets such as maps 
of galaxy clusters,  which contain only isotropic features~\citep{starck:book98}. 
Our isotropic transform is obtained using a scaling function $\phi_{l_c}(\vartheta, \varphi)$ 
with cut-off frequency  $l_c$ and  azimuthal symmetry, meaning that $\phi_{l_c}$ does not depend 
on the azimuth $\varphi$. Hence the spherical harmonic coefficients $\hat \phi_{l_c} (l,m)$ of $\phi_{l_c}$ vanish 
when $m \ne 0$ so that :
\begin{eqnarray}
\phi_{l_c}(\vartheta, \varphi)= \phi_{l_c}(\vartheta) = \sum_{l = 0}^{l = l_c} \hat  \phi_{l_c} (l,0) Y_{l,0}(\vartheta, \varphi)
\end{eqnarray}
where the $Y_{l,m}$ are the spherical harmonic basis functions. Then, convolving a map $f(\vartheta, \varphi)$ with $\phi_{l_c}$ is greatly simplified 
and the spherical harmonic coefficients $\hat c_{0}(l,m)$  of the resulting map $c_0(\vartheta, \varphi)$ are readily given by \citep{bogdanova}:
\begin{eqnarray}
 \hat c_{0}(l,m) = \widehat{\phi_{l_c} * f} (l,m) =  \sqrt{\frac{4\pi}{2l+1} }  \hat \phi_{l_c} (l,0) \hat f(l,m) 
\end{eqnarray}
where $*$ stands for convolution.

\subsubsection*{Multiresolution decompostion}

A sequence of smoother approximations of $f$ on 
a dyadic resolution scale can be obtained using the scaling function $\phi_{l_c}$ as follows
\begin{eqnarray}
c_0   & = &  \phi_{ l_{c} }  * f    \nonumber    \\
c_1   & = &  \phi_{2^{-1} l_{c} }   * f    \nonumber	   \\
&\ldots&\nonumber\\ 
c_j    &=&   \phi_{2^{-j}  l_{c}  }  * f  \nonumber    \\
\end{eqnarray}
where $\phi_{2^{-j} l_{c} }$ is a rescaled version of $\phi_{l_{c}}$ with cut-off frequency $2^{-j} l_{c}$. The above multi-resolution sequence can also  be obtained recursively. 
Define a low pass filter $h_{j}$ for each scale $j$  by 
\begin{eqnarray}
 \hat{H}_{j}(l,m)  =  \sqrt{\frac{4\pi}{2l+1} }  \hat h_{j}(l,m) = \left\{
  \begin{array}{ll}
  \frac {   \hat \phi_{\frac{l_{c}}{2^{j+1}} }(l,m)   }   {  \hat  \phi_{  \frac{l_{c}}{2^{j}} }(l,m)   } & \mbox{if }  l  < \frac{ l_{c}} {2^{j+1}} \quad \textrm{and}\quad m = 0\\
0 & \mbox{otherwise } \ 
  \end{array}
  \right.
\end{eqnarray}
It is then easily shown that $c_{j+1}$ derives from $c_j$ by convolution with $h_j$:  $c_{j+1} = c_{j} * h_j$.

\subsubsection*{The wavelet coefficients}

Given an axisymmetric  wavelet function $\psi_{l_c}$, we can derive in the same way a high pass  filter $g_j$ on each scale $j$:
\begin{eqnarray}
 \hat{G}_{j}(l,m) = \sqrt{\frac{4\pi}{2l+1} }  \hat{g}_{j}(l,m) = \left\{
  \begin{array}{ll}
 \frac {   \hat \psi_{\frac{l_{c}}{2^{j+1}} }(l,m)   }   {  \hat  \phi_{  \frac{l_{c}}{2^{j}} }(l,m)   } & \mbox{if }  l  < \frac{ l_{c}} {2^{j+1}} \quad \textrm{and}\quad m = 0\\ 
1 &\mbox{if }  l  \ge \frac{ l_{c}} {2^{j+1}} \quad \textrm{and}\quad m = 0\\ 
0&  \mbox{otherwise }\
  \end{array}
  \right.
\end{eqnarray}
Using these,  the wavelet coefficients $w_{j+1} $ at scale $j+1$ are obtained from the previous resolution by a simple convolution: $w_{j+1} = c_{j} * g_j$.\\

Just as with the \emph{\`a trous} algorithm, the wavelet coefficients can be defined as the difference between two consecutive resolutions, $w_{j+1}(\vartheta, \varphi) = c_{j}(\vartheta, \varphi) - c_{j+1}(\vartheta, \varphi)$, 
which in fact corresponds to making the following specific choice for the wavelet function $\psi_{l_c}$:
\begin{eqnarray}
\hat \psi_{\frac{l_c}{2^{j}}}(l,m) = \hat \phi_{\frac{l_c}{2^{j-1}}} (l,m)  - \hat \phi_{\frac{l_c}{2^{j}}}(l,m)
\end{eqnarray}
The high pass filters $g_j$ defined above are,  in this particular case,  expressed as: 
\begin{eqnarray}
 \hat{G}_{j}(l,m)  = \sqrt{\frac{4\pi}{2l+1} } \hat{g}_{j}(l,m) =  1 - \sqrt{\frac{4\pi}{2l+1} } \hat{h}_j(l,m)   =   1 - \hat{H}_j(l,m) 
\end{eqnarray}
Obviously, other wavelet functions could be used as well.

\subsubsection*{Choice of the scaling function}
Any function with a cut-off frequency is a possible candidate. We retained here 
a B-spline function of order 3. It is very close to a Gaussian function and converges
rapidly to $0$:
\begin{eqnarray}
\hat \phi_{l_c} (l,m = 0) =\frac{3}{2} B_{3}( \frac{2 l}{l_{c} })  
\end{eqnarray}
where $B(x) = \frac{1}{12}({\mid{x-2}\mid}^3 - 4 {\mid{x-1}\mid}^3 + 6 {\mid{x}\mid}^3 - 4 {\mid{x+1}\mid}^3 + {\mid{x+2}\mid}^3)$.

\begin{figure*}[htb]
\centerline{
\hbox{
}}
\caption{On the left, the scaling function $\hat{\phi}$ and, on the 
right, the wavelet function $\hat{\psi}$.}
\label{fig_diff_uv_phi_psi}
\end{figure*}

In Fig.~\ref{fig_diff_uv_phi_psi} the chosen scaling function 
derived from a $B$-spline of degree 
3, and its resulting wavelet function, are plotted in frequency space.

\begin{center}
\begin{tabular}{|c|} \hline
\begin{minipage}[b]{5.3in}
\vspace{0.1in}

\small{
\textsf{1. Compute the $B_3$-spline scaling function and derive $\psi$, $h$ and $g$ numerically.}

\textsf{2. Compute the corresponding Spherical Harmonics of image $c_0$. We get ${\hat c}_0$.}

\textsf{3. Set $j$ to $0$. Iterate:}

\hspace{0.3in} \textsf{4. Multiply  $\hat{c}_j$ by $\widehat H_{j}$. We get the array $\hat{c}_{j+1}$.}

\hspace{0.33in} \textsf{Its inverse Spherical Harmonics Transform gives the image at scale $j+1$.}

\hspace{0.3in} \textsf{5. Multiply $\hat{c}_j$ by $\widehat G_{j}$. We get the complex array
$\hat{w}_{j+1}$.}

\hspace{0.33in} \textsf{ The inverse Spherical Harmonics transform of $\hat{w}_{j+1}$ 
gives the wavelet coefficients $w_{j+1}$ at scale $j+1$.}

\hspace{0.3in} \textsf{6. j=j+1 and if $j \leq  J$, return to Step 4.}

\textsf{7. The set $\{w_1, w_2, \dots, w_{J}, c_{J}\}$ describes the wavelet transform on the sphere of $c_0$.}}

\vspace{0.05in}
\end{minipage}
\\\hline
\end{tabular}
\\ \vspace{0.1in}
The numerical algorithm for the undecimated wavelet transform on the sphere.
\end{center}
\linespread{1.3}

\subsubsection*{Reconstruction}

If the wavelet is the difference between two resolutions, Step 5 in the above UWTS algorithm can be replaced by the following simple subtraction $w_{j+1} = c_{j} - c_{j+1}$. In this case, the reconstruction of an image from its wavelet coefficients ${\cal W} = \{w_1,\dots, w_{J}, c_{J}\}$ 
is straightforward: 
\begin{eqnarray}
 c_{0}(\theta, \phi) = c_{J}(\theta, \phi) + \sum_{j=1}^J  w_j(\theta, \phi)
\end{eqnarray}
This is the same reconstruction formula as in the \emph{\`a trous} algorithm: the simple 
sum of all scales reproduces the original data. However, since the present decomposition is redundant, the procedure for reconstructing an image from its coefficients is not unique and this can profitably be used to impose additional constraints on the synthesis functions (\emph{e.g.} smoothness, positivity) used in the reconstruction. Here for instance, using the relations:
\begin{eqnarray}
\hat c_{j+1}(l,m) = \widehat H_{j} (l,m)  \hat c_{j} (l,m) \nonumber \\
\hat w_{j+1}(l,m) = \widehat G_{j} (l,m) \hat c_{j} (l,m) 
\end{eqnarray}
a least squares estimate of $c_j$ from $c_{j+1}$ and $w_{j+1}$ gives:
\begin{eqnarray}
\hat{c}_{j}   = \hat{c}_{j+1}  {\widehat {\tilde H}}_{j}   + \hat{w}_{j+1}  {\widehat {\tilde G}}_{j} 
\end{eqnarray}
where the conjugate filters $ {\widehat {\tilde H}}_j $ and $ {\widehat {\tilde G}}_j$ have the expression:
\begin{eqnarray}
 {\widehat {\tilde H}}_j =  \sqrt{\frac{4\pi}{2l+1} } {\hat {\tilde h}}_j & = {\widehat H}_{j}^* /
(\mid {\widehat H}_{j}\mid^2 + \mid {\widehat G}_j\mid^2) \label{eqnht} \\ 
 {\widehat {\tilde G}}_j =  \sqrt{\frac{4\pi}{2l+1} } {\hat {\tilde g}}_j & = {\widehat G}_{j}^* /
(\mid {\widehat H}_j \mid^2 + \mid {\widehat G}_j \mid^2)
\label{eqngt}
\end{eqnarray}
and the reconstruction algorithm is:
\begin{center}
\begin{tabular}{|c|} \hline
\begin{minipage}[b]{5.3in}
\vspace{0.1in}

\small{
\textsf{1. Compute the $B_3$-spline scaling function and derive $\hat \psi$, $\hat h$, $\hat g$, ${\hat {\tilde h}}$, ${\hat {\tilde g}}$ numerically.}

\textsf{2. Compute the corresponding Spherical Harmonics of the image at the low resolution $c_J$. We get ${\hat c}_J$.}

\textsf{3. Set $j$ to $J-1$. Iterate:}

\hspace{0.3in} \textsf{4. Compute the Spherical Harmonics transform 
  of the wavelet coefficients $w_{j+1}$ at scale $j+1$. We get  $\hat{w}_{j+1}$.}

\hspace{0.3in} \textsf{5. Multiply  $\hat{c}_{j+1}$ by ${\widehat {\tilde H}}_j $.}

\hspace{0.3in} \textsf{6. Multiply $\hat{w}_{j+1}$ by $ {\widehat {\tilde G}}_j $.}

\hspace{0.3in} \textsf{7. Add the results of steps 6 and 7. We get $\hat  c_j$.}

\hspace{0.3in} \textsf{8. j=j-1 and if $j \ge  0$, return to Step 4.}

\textsf{9. Compute The inverse Spherical Harmonic transform of $\hat  c_0$}}

\vspace{0.05in}
\end{minipage}
\\\hline
\end{tabular}
\\ \vspace{0.1in}
\end{center}
\linespread{1.3}

The synthesis low pass and high pass filters $\hat{\tilde h}$ and $\hat{\tilde g}$ are 
plotted in Fig.~\ref{fig_diff_uv_ht_gt}. 

\begin{figure*}[htb]
\centerline{
\hbox{
}}
\caption{On the left, the filter $\hat{\tilde{h}}$, and on the right the 
filter $\hat{\tilde{g}}$.}
\label{fig_diff_uv_ht_gt}
\end{figure*}

\begin{figure*}
\vbox{
\centerline{
\hbox{
}}
\centerline{
\hbox{
}}
\centerline{
\hbox{
}}
}
\caption{{ WMAP data and its wavelet transform on the sphere using five resolution levels 
 (4 wavelet scales and the coarse scale). The sum of these five maps reproduces exactly the original data (top left).
 Top: original data and the first wavelet scale. Middle:  the second and third wavelet scales. 
 Bottom:  the fourth wavelet scale and the last smoothed array.}}
\label{Figure:UWTS}
\end{figure*}

\begin{figure*}
\centerline{
\hbox{
\psfig{figure=fig_backwt_sphere.ps,bbllx=0.5cm,bblly=6.5cm,bburx=20.5cm,bbury=21.5cm,height=7.5cm,width=10cm,clip=}
}}
\caption{{ Backprojection of a wavelet coefficient at different scales. Each map is obtained by setting all but one of the wavelet
coefficients  to zero, and applying an inverse wavelet transform. Depending on the scale and the position
of the non zero wavelet coefficient, the reconstructed image presents an isotropic feature with a given size.}}
\label{Figure:back_wt}
\end{figure*}

{ Figure~\ref{Figure:UWTS} shows the WMAP data (top left) and 
its undecimated wavelet decomposition on the sphere using five resolution levels.
Figures~\ref{Figure:UWTS} middle left, middle right and bottom left show respectively 
the four wavelet scales. Figure~\ref{Figure:UWTS} bottom right shows the last smoothed array.
Figure~\ref{Figure:back_wt} shows the backprojection of a wavelet coefficient at different scales
and positions.}

\subsection{Isotropic Pyramidal Wavelet Transform on the Sphere (PWTS) }

\begin{figure*}
\vbox{
\centerline{
\hbox{
}}
\centerline{
\hbox{
}}
\centerline{
\hbox{
}}
}
\caption{{ WMAP data (top left) and its pyramidal wavelet transform on the sphere using five resolution levels 
 (4 wavelet scales and the coarse scale). The original map can be reconstructed exactly from
 the pyramidal wavelet coefficients.
 Top: original data and the first wavelet scale. Middle:  the second and third wavelet scales. 
 Bottom:  the fourth wavelet scale and the last smoothed array. The number of pixels is divided by four
 at each resolution level, which can be helpful when the data set is large.}}
\label{Figure:PWTS}
\end{figure*}

In the previous algorithm, no downsampling is
performed and each scale of the wavelet decomposition has the same number of pixels 
as the original data set. Therefore, the number of pixels in the
decomposition is equal to the number of pixels in the data multiplied by the number of scales.
For applications such as PLANCK data restoration, we may prefer to introduce some decimation
in the decomposition so as to reduce the required memory size and the computation time.
This can be done easily by using a specific property of the chosen scaling function.
Indeed, since we are considering here a scaling function with an initial cut-off $l_c$ in spherical harmonic multipole number $l$, and since the actual cut-off is reduced by a factor of two at each step, the number 
of significant spherical harmonic coefficients is then reduced by a factor of four after each convolution with the low pass filter
$h$. Therefore, we need fewer pixels in the direct space when we compute the inverse 
spherical harmonic transform.  Using the Healpix pixelization scheme \citep{healpix},  
this can be done easily by dividing by 2 the 
{\it nside} parameter when resorting to the inverse spherical harmonic transform routine.

{ Figure~\ref{Figure:PWTS} shows WMAP data (top left) and its pyramidal wavelet transform using five scales.
As the scale number increases (i.e. the resolution decreases), the pixel size becomes
larger. Figures~\ref{Figure:PWTS} top right, middle left, middle right and bottom left show respectively 
the four wavelet scales. Figure~\ref{Figure:PWTS} bottom right shows the last smoothed array.}

\section{Curvelet Transform on the Sphere (CTS) }
\label{sect_cur}
\subsection{Introduction.}

The 2D curvelet transform, proposed in \citet{cur:donoho99}, \citet{starck:sta01_3} and \citet{starck:sta02_3} enables the
directional analysis of an image in different scales. 
The fundamental property of the curvelet transform is to 
analyze the data with functions of length of about
 $2^{-j/2}$ for the $j^{\textrm{th}}$ sub-band $[2^j, 2^{j+1}]$ of the 
two dimensional wavelet transform.   
Following the implementation described in \citet{starck:sta01_3} and \citet{starck:sta02_3}, the data first undergoes an  Isotropic Undecimated Wavelet Transform (i.e. \emph{\`a trous} algorithm). Each scale $j$ is then decomposed into smoothly overlapping blocks of
side-length $B_j$ pixels in such a way that the overlap between two
vertically adjacent blocks is a rectangular array of size $B_j \times B_j/2$.
Finally, the ridgelet transform \citep{cur:candes99_1}  is applied on each individual block which amounts  to applying  a 1-dimensional wavelet transform to the slices of its Radon transform. More details on the implementation of the digital curvelet transform
can be found in Starck et al\citep*{starck:sta01_3,starck:sta02_3}.
It has been shown that the curvelet transform could be very useful for the 
detection and the discrimination of sources of  non-Gaussianity in CMB \citep{starck:sta02_4}.
The curvelet transform is also redundant, with a redundancy factor of $16J+1$ whenever $J$ scales
are employed. Its complexity scales like that of the ridgelet transform that is  as $O(n^2 \log_2n)$. This method is best for the detection of anisotropic structures and smooth curves and edges of different lengths.

\subsection{Curvelets on the Sphere.}
The Curvelet transform on the sphere (CTS) can be similar to the 2D digital curvelet transform,
but replacing the \`a trous algorithm by the Isotropic Wavelet Transform on the Sphere previously 
described. The CTS algorithm consists of the following three steps :
\begin{itemize}
\item {\it Isotropic Wavelet Transform on the Sphere.}  
\item {\it Partitioning.} Each scale is decomposed into blocks of 
an appropriate scale (of side-length $\sim2^{-s}$), using the Healpix pixelization.
\item {\it Ridgelet Analysis.} Each square is analyzed via the discrete ridgelet transform.
\end{itemize}

\subsubsection*{Partitioning using the Healpix representation.}
The Healpix representation is a curvilinear partition of the sphere into quadrilateral pixels of exactly equal area but with varying shape. The base resolution divides the sphere into 12 quadrilateral faces of equal area placed on three rings around the poles and equator.  Each face is subsequently divided into $nside^{2}$ pixels following a hierarchical quadrilateral tree structure. 
The geometry of the Healpix sampling grid makes it easy to partition a spherical map into blocks of a specified size $2^n$. We first extract  the twelve base-resolution faces, and each face is then decomposed into overlapping blocks as in the 2D digital curvelet transform. With this scheme however, there is no overlapping between blocks belonging to different base-resolution faces. This may result in blocking effects for example in denoising experiments \emph{via} non linear filtering. A simple way around this difficulty is to work with various rotations of the data with respect to the sampling grid. 

\subsubsection*{Ridgelet transform}
Once the partitioning is performed, the standard 2D ridgelet transform described in \citet{starck:sta02_3}  is applied  in each individual block :
\begin{enumerate}
\item Compute the 2D Fourier transform.
\item Extract lines going through the origin in the frequency plane.
\item Compute  the 1D inverse Fourier transform of each line. This achieves the Radon transform of the current block.
\item Compute the 1D wavelet transform of the lines of the Radon transform.
\end{enumerate}
The first three steps correspond to a Radon transform method called the {\it linogram}.
Other implementations of the Radon transform,  such as the {\it Slant Stack Radon Transform} \citep{cur:donoho_02},
can be used as well, as long as they offer an exact reconstruction.
   
Figure~\ref{Figure:rid_sphere} shows the flowgraph of the ridgelet transform on the sphere
and Figure~\ref{Figure:back_rid} shows the backprojection of a ridgelet coefficient at 
different scales and orientations.

\begin{figure*}
\centerline{
\hbox{
}}
\caption{Flowgraph of the Ridgelet Transform on the Sphere.}
\label{Figure:rid_sphere}
\end{figure*}

\begin{figure*}
\centerline{
\hbox{
}}
\caption{Backprojection of a ridgelet coefficient at different scales and orientations.
{Each map is obtained by setting all but one of the ridgelet
coefficients to zero, and applying an inverse ridgelet transform. Depending on the scale and the position
of the non zero ridgelet coefficient, the reconstructed image presents a feature with a given width and a given
orientation.}}
\label{Figure:back_rid}
\end{figure*}

\subsection{Algorithm}
The curvelet transform algorithm on the sphere is as follows:
\begin{enumerate}
\item Apply the isotropic wavelet transform on the sphere with $J$ scales,
\item Set the block size $B_1 = B_{min}$,
\item For $j = 1, \ldots, J$ do,
\begin{itemize}
\item partition the subband $w_j$ with a block size $B_j$ and apply the
digital ridgelet transform to each block,
\item if $j \mbox{ modulo } 2 = 1$ then $B_{j+1} = 2 B_{j}$,
\item else $B_{j+1} = B_{j}$.
\end{itemize}
\end{enumerate}
The sidelength of the localizing windows is doubled {\em at every
other} dyadic subband, hence maintaining the fundamental property of
the curvelet transform which says that elements of length about
$2^{-j/2}$ serve for the analysis and synthesis of the $j$-th subband
$[2^j, 2^{j+1}]$.  We used the default value $B_{min} = 16$
pixels in our implementation. Figure~\ref{Figure:cur_sphere}
gives an overview of the organization of the algorithm.
\begin{figure*}
\centerline{
\hbox{
\psfig{figure=fig_flowgraph_curvelet_sphere.ps,bbllx=1cm,bblly=7cm,bburx=20cm,bbury=21cm,height=9cm,width=12cm,clip=}
}}
\caption{Flowgraph of the Curvelet Transform on the Sphere.}
\label{Figure:cur_sphere}
\end{figure*}
\begin{figure*}
\centerline{
\hbox{
}}
\caption{Backprojection of a curvelet coefficient at different scales and orientations.
{ Each map is obtained by setting all but one of the curvelet
coefficients to zero, and applying an inverse curvelet transform. Depending on the scale and the position
of the non zero curvelet coefficient, the reconstructed image presents a feature with a given width, length and
orientation.}}
\label{Figure:back_cur}
\end{figure*}
Figure~\ref{Figure:back_cur} shows the backprojection of  curvelet coefficients at 
different scales and orientations.

\subsection{Pyramidal Curvelet Transform on the Sphere (PCTS)}
The CTS is very redundant, which may be a problem in handling huge data sets such as the future
PLANCK data. The redundancy can be reduced by substituting,  in the 
curvelet transform algorithm, the pyramidal wavelet transform to the undecimated wavelet transform.
The second step which consists in applying the ridgelet transform on the wavelet scale is unchanged.
The pyramidal curvelet transform algorithm is:
\begin{enumerate}
\item Apply the pyramidal wavelet transform on the sphere with $J$ scales,
\item Set the block size $B_1 = B_{min}$,
\item For $j = 1, \ldots, J$ do,
\begin{itemize}
\item partition the subband $w_j$ with a block size $B_j$ and apply the
digital ridgelet transform to each block,
\item if $j \mbox{ modulo } 2 = 2$ then $B_{j+1} = B_{j} / 2$,
\item else $B_{j+1} = B_{j}$.
\end{itemize}
\end{enumerate}
In the next section, we show how the pyramidal curvelet transform can be used for image filtering.

\section{Filtering}
\label{sect_exp}
\subsection{Hard thresholding}
Wavelets and Curvelets have been used successfully  in image 
denoising  \emph{via} non-linear filtering or thresholding methods \citep{starck:book02,starck:sta01_3}. Hard thresholding, for instance, consists in setting all insignificant coefficients (\emph{i.e.} coefficients with an absolute value below a given threshold)
to zero. In practice, we need to estimate the noise standard deviation $\sigma_j$ in each band $j$
and a wavelet (or curvelet) coefficient $w_j$ is significant if $\mid w_j \mid > k \sigma_j$,
where $k$ is a user-defined parameter, typically chosen between 3 and 5. 
The $\sigma_j$ estimation in band $j$ can be derived from simulations \citep{starck:book02}. 
Denoting as $D$ the noisy data and $\delta$ the thresholding operator, the filtered data $\tilde D$ are obtained by : 
\begin{eqnarray}
 {\tilde D} =    {\cal R} \delta( {\cal T} D)
\end{eqnarray}
where ${\cal T}$ is the wavelet (resp. curvelet) transform operator and ${\cal R}$ is 
the wavelet (resp. curvelet) reconstruction operator. 

\begin{figure*}
\vbox{
\centerline{
\hbox{
}}
\centerline{
\hbox{
}}
\centerline{
\hbox{
}}
}
\caption{\textbf{Denoising.} Upper left and right : simulated synchrotron image and same image with
an additive Gaussian noise (\emph{i.e.} simulated data).  Middle: pyramidal wavelet filtering and residual.
Bottom: pyramidal curvelet filtering  and residual. { On such data, displaying very anisotropic features, 
the residual with curvelet denoising  is cleaner than with the wavelet denoising.}}
\label{Figure:sync_filter}
\end{figure*}

Figure~\ref{Figure:sync_filter} describes the setting and the results of a simulated denoising experiment : 
upper left, the original simulated map of the  synchrotron emission (renormalized between 0 and 255); 
upper right, the same image plus additive Gaussian noise ($\sigma=5$); middle, the pyramidal wavelet 
filtered image and the residual (i.e. noisy data minus filtered data); bottom, the 
pyramidal curvelet transform filtered image and the residual. A $5 \sigma_j$ detection threshold 
was used in both cases. On such data, displaying very anisotropic features, using curvelets produces 
better results than the wavelets.

\subsection{The Combined Filtering Method on the Sphere}

{\small
\begin{table*}[htb]
\baselineskip=0.4cm
\begin{center}
\begin{tabular}{lccccc} \hline \hline
Method                          &  Error Standard Deviation     &  SNR (dB)    \\ \hline \hline
Noisy map                       & 5.  &      13.65  \\
Wavelet                         & 1.30  &    25.29  \\
Curvelet                        & 1.01  &    27.60  \\
CFM                             & 0.86  &    28.99  \\ \hline
\hline
\end{tabular}
\caption{Table of error standard deviations and SNR values after filtering the synchrotron noisy map
        (Gaussian white noise - sigma = 5 ) by the wavelet, the curvelet and the combined filtering method.
Images are available at "http://jstarck.free.fr/mrs.html".}
\label{comptab_sync}
\end{center}
\end{table*}
}

\begin{figure*}
\centerline{
\hbox{
}}
\caption{Denoising. Combined Filtering Method (pyramidal wavelet and pyramidal curvelet) and residual.}
\label{Figure:sync_cbf_filter}
\end{figure*}

\begin{figure*}
\centerline{
\hbox{
}}
\caption{{\bf Combined Filtering Method, face 6 in the Healpix representation of the image 
shown in Figure~\ref{Figure:sync_cbf_filter}. 
From top to bottom and left to right, respectively the a) original image face, b) the noisy image,
c) the combined filtered image, d) the combined filtering residual, e) the wavelet filtering residual and
f) the curvelet filtering residual.}}
\label{Figure:sync_face_cbf_filter}
\end{figure*}

The residual images for both the
wavelet and curvelet transforms shown in Figure~\ref{Figure:sync_filter}
show that wavelets do not restore long features with high fidelity while curvelets
are seriously challenged by isotropic or small features. Each transform has its own area of expertise 
and this complementarity is of great potential. The Combined Filtering Method (CFM) \citep{starck:spie01a}
allows us to benefit from the advantages of both transforms. This iterative method detects the significant coefficients
in both the wavelet domain and the curvelet domain and guarantees  that the reconstructed map will take into
account any pattern detected as significant by either of the transforms. 
A full description of the algorithm is given in Appendix 1.
Figure~\ref{Figure:sync_cbf_filter} shows the CFM denoised image and its residual.
{ Figure~\ref{Figure:sync_face_cbf_filter} shows one face (face 6) of the following Healpix images:
upper left, original image; upper right, noisy image; middle left, 
restored image after denoising by the combined transform; middle 
right, the residual; bottom left and right, the residual using respectively the curvelet and the 
wavelet denoising method.  }
The results are reported in Table~\ref{comptab_sync}. 
The residual is much better when the combined filtering is applied, and no
feature can be detected any longer by eye in the residual.
Neither with the wavelet filtering nor with the curvelet filtering could such a clean residual be obtained.


\section{Wavelets on the Sphere and Independent Component Analysis}
\label{sect_ica}
\subsection{Introduction}
 Blind Source Separation (BSS) is a problem that occurs in multi-dimensional data processing. 
 The goal is to recover unobserved signals, images or \emph{sources} $S$ from 
 mixtures $X$ of these sources observed typically at the output of an array of $m$ sensors. 
 A simple mixture model would be linear and instantaneous with additive noise as in:
\begin{equation}\label{model0}
X = A S + N
\end{equation}
where $X$, $S$ and $N$ are random vectors of respective sizes $m \times 1$, $n\times 1$ and $m \times 1$, and $A$ is an $m\times n$ matrix.
Multiplying $S$ by $A$ linearly mixes the $n$ sources resulting in $m$ observed mixture processes corrupted by  additive instrumental Gaussian noise $N$. 
Independent Component Analysis methods were developed to solve the BSS problem, \emph{i.e.} 
given a batch of $T$ observed samples of $X$, estimate $A$ and $S$, relying mostly on the statistical independence of the source 
processes.

Algorithms for blind component separation and mixing
matrix estimation depend on the \emph{a priori} model used for the probability
distributions of the sources~\citep{ica:3easy} although rather coarse assumptions can be made \citep{ica:tutorial,ica:icabook}. 
 In a first set of techniques,
source separation is achieved in a noise-less setting, based on the 
non-Gaussianity of all but possibly one of the components. 
Most mainstream ICA techniques belong to this category : JADE~\citep{ica:JADE}, 
FastICA, Infomax~\citep{ica:icabook}. In a second set of blind techniques, the components 
are modeled as Gaussian processes and, 
in a given representation (Fourier, wavelet, etc.), separation requires that the sources 
have diverse, \emph{i.e.} non proportional, variance profiles. 
The Spectral Matching ICA method (SMICA) \citep{ica:Del2003,starck:yassir05} considers in this sense the case of mixed stationary Gaussian components in a noisy context : 
moving to a Fourier representation, the point is that colored 
components can be separated based on the diversity of their power spectra. 

\subsection{SMICA: from Fourier to Wavelets}

 SMICA, which was designed to address some of the general problems raised 
 by Cosmic Microwave Background data analysis, has already
shown significant success for CMB spectral estimation in multidetector
experiments~\citep{ica:Del2003,ica:patanchon}. However, SMICA suffers from 
 the non locality of the Fourier transform which has undesired effects when dealing 
 with non-stationary components or noise, or with incomplete data maps. 
 The latter is a common issue in astrophysical data analysis:  
 either the instrument scanned only a fraction of the sky or some regions of 
 the sky  were masked due to localized strong
astrophysical sources of contamination ( compact radio-sources or
galaxies, strong emitting regions in the galactic plane). A simple way to 
overcome these effects is to move instead to a wavelet representation so as 
to benefit from the localization property of  wavelet filters,
which leads to wSMICA~\citep{starck:yassir05}.
The wSMICA method uses an undecimated \emph{\`a trous} algorithm 
with the cubic box-spline~\citep{starck:book02} as the
scaling function. This transform has several favorable properties 
for astrophysical data analysis. In particular, it is a shift invariant transform, 
the wavelet coefficient maps
on each scale are the same size as the initial image, and the wavelet
and scaling functions have small compact supports in the initial representation.
All of these allow  missing patches in the data maps to be handled easily.

Using this wavelet transform algorithm, the multichannel data $X$ is decomposed 
into $J$ detail maps  $X_{j}^w$ and a smooth approximation map $X_{J+1}^w$ 
over a dyadic resolution scale.
Since applying a wavelet transform on (\ref{model0}) does not 
affect the mixing  matrix $A$, the covariance matrix of the observations 
at scale $j$ is still structured as  
\begin{equation}\label{structure_w}
R_w^X(j) = A R_w^S(j) A^{\dagger} +   R_w^N(j)
\end{equation}
where  $R_w^S(j)$ and  $R_w^N(j)$ are the diagonal spectral covariance
matrices in the wavelet representation of $S$ and $N$ respectively.
It was shown \citep{starck:yassir05} that replacing in the SMICA method
the covariance matrices  derived from the Fourier coefficients by the 
covariance matrices derived from wavelet coefficients leads to much better
results when the data are incomplete. 
This is due to the fact that the wavelet filter response on scale $j$  
is short enough compared to data size and gap widths  and most of the samples in the filtered
signal then remain unaffected by the presence of gaps.  
Using these samples exclusively yields an estimated covariance matrix $\widehat{R}_w^X(j)$ 
which is not biased by the missing data. 

\subsection{SMICA on the Sphere: from spherical harmonics to wavelets on the sphere (wSMICA-S)}

Extending SMICA to deal with multichannel data mapped to the sphere is 
straightforward~\citep{ica:Del2003}. The idea is simply to substitute 
the spherical harmonics transform for the Fourier transform used in the 
SMICA method. Then, data covariance matrices are estimated 
in this representation over intervals in multipole number $l$. 
However SMICA on spherical maps still suffers from the non locality of the spherical 
harmonics transform whenever the data to be processed is non stationary, 
incomplete, or partly masked. 

Moving to a wavelet representation allows one to overcome these effects : 
wSMICA can be easily implemented for data mapped to the sphere by substituting 
the UWTS described in section~\ref{sect_wts} to the undecimated \emph{\`a trous} algorithm 
used in the previous section. In the linear mixture model~(\ref{model0}),  
$X$ now stands for an array of observed spherical maps,  
$S$ is now an array of spherical source maps to be recovered  and $N$ is an array 
of spherical noise maps. The mixing matrix $A$ achieves a pixelwise 
linear mixing of the source maps.  Applying the above UWTS on both 
sides of~(\ref{model0}) does not affect the mixing matrix $A$ so that,  
assuming independent source and noise processes, the covariance matrix of the 
observations at scale $j$ is again structured as in~(\ref{structure_w}).  
Source separation follows from  minimizing the  covariance matching 
criterion~(\ref{Cost_wavelet}) in this \emph{spherical wavelet} representation.
A full description is given in Appendix~2.

\subsection{Experiments}
\label{sec:NUMEXP}
As an application of wSMICA on the sphere, we consider here the problem of CMB data analysis  but in the special case where the use of a galactic mask is a cause of non-stationarity which impairs the use of the spherical harmonics transform. 

The simulated CMB, galactic dust and Sunyaev Zel'dovich (SZ) maps used, shown on the left-hand side of Figure~\ref{components}, were obtained as described in~\citet{ica:Del2003}.  The problem of  instrumental point spread functions is not addressed here, and all maps are assumed to have the same resolution. The high level foreground emission from the galactic plane region was removed using the $Kp2$ mask from the WMAP team website\footnote{\emph{http://lambda.gsfc.nasa.gov/product/map/intensity\_mask.cfm}}.  These three \emph{incomplete} maps were mixed using the matrix in Table~\ref{MatrixA},  in order  to simulate observations in the six channels of the Planck high frequency instrument~(HFI).

\begin{table}[!h]
   \footnotesize {
    \begin{tabular}{@{} cccc|c @{}}
      CMB & DUST & SZ &  & channel \\
      & & & &\\
      \hline
      & & & &\\
      $\quad1.0\quad$  &  $ 1.0 $    		&  $\quad-1.51\quad$	 	&& 100~GHz\\
      $\quad1.0 \quad$ & $ 2.20$   		& $\quad-1.05\quad$  		&& 143~GHz\\
      $\quad1.0 \quad$ &  $ 7.16 $  		& $\quad0.0\quad $ 			&& 217~GHz\\
      $\quad1.0 \quad$ &   $56.96 $		&  $\quad 2.22\quad$ 	&&353~GHz\\
      $\quad1.0 \quad$ &  $1.1\times10^{3}$ 		&   $\quad5.56\quad$  	&&545~GHz\\
      $\quad1.0\quad $ &  $1.47\times10^{5}$  	& $ \quad11.03\quad $	&& 857~GHz\\
    \end{tabular}}
    \caption{Entries of $A$, the mixing matrix used in our simulations.}\label{MatrixA}
\end{table}

Gaussian \emph{instrumental} noise was added in each channel according to model~(\ref{model0}).
The relative noise standard deviations between channels were set
according to the nominal values of the Planck HFI given in Table~\ref{NoiseScale}. 
{\small
\begin{table}[htb]
     \footnotesize {
    \begin{tabular}{@{} cccccc|c @{}}
      100&143&217&353&545&857&channel \\
          & & & & & &\\
  \hline
          & & & & & &\\
    $ 2.65\!\!\times\!\!10^{-6}$&$2.33\!\!\times\!\!10^{-6}$&$3.44\!\!\times\!\!10^{-6}$&$1.05\!\!\times\!\!10^{-5}$&$1.07\!\!\times\!\!10^{-4}$& $4.84\!\!\times\!\!10^{-3}$ &noise std\\
          & & & & & &\\
      \end{tabular}}
    \caption{Nominal noise standard deviations in the six channels of the Planck HFI.}\label{NoiseScale}
\end{table}
}

The synthetic observations were decomposed into six scales using the isotropic UWTS  and  wSMICA was used to obtain estimates of the mixing matrix and  of  the initial source templates. The resulting component maps estimated using wSMICA, for nominal noise levels, are shown on the right-hand side of Figure~\ref{components} where the quality of the reconstruction can be visually assessed by comparison to the initial components. The component separation was also performed with SMICA based on Fourier statistics computed in  the same six  dyadic bands imposed by our choice of wavelet
transform, and with JADE.  In Figure~\ref{resultats}, the performances of SMICA, wSMICA and JADE are compared in the particular case of CMB map estimation, in terms of the relative standard deviation of the reconstruction error, $MQE$,  defined by
 \begin{equation}
  MQE  = \frac{\mathbf{std}  ( CMB(\vartheta, \varphi)  - \alpha \times \widehat{CMB}(\vartheta, \varphi)  )}{\mathbf{std}  ( CMB(\vartheta, \varphi)   ) }
  \label{MQE}
\end{equation}  
where $\mathbf{std}$ stands for empirical standard deviation ( obviously computed outside the masked regions), and $\alpha$ is a linear regression coefficient estimated in the least squares sense.  As expected, since it is meant to be used in a noiseless setting,  JADE performs well when the noise is very low. However, as the noise level increases, its performance degrades quite rapidly compared to the covariance matching methods.  Further, these results clearly show that using wavelet-based covariance matrices provides a simple and
efficient way to treat the effects of gaps on the
performance of source separation using statistics based on the non local Fourier representation. 

\begin{figure*}[htb]
\vbox{
\centerline{
\hbox{
}}
\centerline{
\hbox{
}}
\centerline{
\hbox{
}}}
\caption{\textbf{Left :} zero mean templates for CMB ($\sigma =  4.17\times 10^{-5}$), 
galactic dust ($\sigma =  8.61\times 10^{-6}$) and SZ ($\sigma =  3.32\times 10^{-6}$) used in 
the experiment described in section~\ref{sec:NUMEXP}. The standard deviations given are for 
the region outside the galactic mask.  The maps on the right were estimated using wSMICA and 
scalewise Wiener filtering { as is explained in Appendix~2}. Map reconstruction using 
Wiener filtering clearly is optimal only in front of
stationary Gaussian processes.  For non Gaussian maps, such as given by the Sunyaev Zel'dovich  
effect,  better reconstruction can be expected from 
non linear methods.  The different maps are drawn here in different color  scales  
to enhance structures and ease visual comparisons.  }
\label{components}
\end{figure*}

\begin{figure*}
\begin{center}
\caption{Relative reconstruction error defined by~(\ref{MQE}) of the CMB component map using SMICA, 
wSMICA and JADE as a function of the instrumental noise level in dB relative to the 
nominal values in table~\ref{NoiseScale}. { The zero dB noise level corresponds to the expected noise
in the future PLANCK mission}.} 
\label{resultats}
\end{center}
\end{figure*}

\section{Conclusion}

We have introduced new multiscale decompositions on the sphere, the wavelet transform, 
the ridgelet transform and
the curvelet transform.
It was shown in \citet{starck:sta03_1} that combining the
statistics of wavelet coefficients and curvelet coefficients of flat maps leads to a powerful analysis 
of the non-Gaussianity
in the CMB. Using the new decompositions proposed here, it is now possible to perform such  
a study on data mapped to the sphere such as  from the WMAP or PLANCK experiments.  
For the denoising application, we have shown that the Combined Filtering Method allows us to significantly  improve 
the result compared to a standard hard thresholding in a given transformation.
Using the isotropic UWTS and statistics in this representation also allows us to properly treat incomplete or non-stationary data on the sphere which cannot be done in the non-local Fourier representation. This is clearly illustrated by the results obtained with wSMICA which achieves component separation in the wavelet domain thus reducing the negative impact that gaps have on the performance of source separation using the initial SMICA algorithm in the spherical harmonics representation. 
Further results are available at
\begin{verbatim}
       http://jstarck.free.fr/mrs.hmtl
\end{verbatim}
as well as information about the IDL code for the described transformations based on the Healpix package \citep{healpix}.

\begin{acknowledgements}
We wish to thank David Donoho, Jacques Delabrouille, 
Jean-Fran\c{c}ois Cardoso and Vicent Martinez for useful discussions. 

\end{acknowledgements}


\section*{Appendix 1: The Combined Filtering Method}

In general, suppose that we are given $K$ linear transforms $T_1,
\ldots, T_K$ and let $\alpha_k$ be the coefficient sequence of an
object $x$ after applying the transform $T_k$, i.e. $\alpha_k = T_k
x$. We will assume that for each transform $T_k$,  a
reconstruction rule is available that we will denote by $T^{-1}_k$, although this is
clearly an abuse of notation. $T$ will denote the block
diagonal matrix with $T_k$ as building blocks and $\alpha$ the
amalgamation of the $\alpha_k$.

A hard thresholding rule associated with the transform $T_k$ synthesizes 
an estimate $\tilde{s}_k$ via the formula 
\begin{equation}
\label{eq:ht}
\tilde{s}_k = T_k^{-1} \delta(\alpha_k)
\end{equation}
where $\delta$ is a rule that sets to zero all the coordinates of
$\alpha_k$ whose absolute value falls below a given sequence of
thresholds (such coordinates are said to be non-significant).
 
Given data $y$ of the form $y = s + \sigma z$, where $s$ is the image
we wish to recover and $z$ is standard white noise, we propose to solve
the following optimization problem \citep{starck:spie01a}:
\begin{equation}
  \label{eq:l1-min}
  \min \|T\tilde{s}\|_{\ell_1}, \quad \mbox{subject to} \quad s \in C,  
\end{equation}
where $C$ is the set of vectors $\tilde{s}$ 
that obey the linear constraints
\begin{equation}
\label{eq:constraints}
\left\{  \begin{array}{ll}
  \tilde{s} \ge 0, \\
  |T\tilde{s} - Ty| \le e; 
  \end{array}
  \right. 
\end{equation}
here, the second inequality constraint 
only concerns the set of significant coefficients, 
i.e. those indices $\mu$ such that $\alpha_\mu =
(Ty)_\mu$ exceeds (in absolute value) a threshold $t_\mu$. Given a
vector of tolerance $(e_\mu)$, we seek a solution whose coefficients
  $(T\tilde{s})_\mu$ are within $e_\mu$ of the noisy
empirical $\alpha_\mu$.  Think of $\alpha_\mu$ as being given by
\[
y = \langle y, \varphi_\mu \rangle, 
\]
so that $\alpha_\mu$ is normally distributed with mean $\langle f,
\varphi_\mu \rangle$ and variance $\sigma^2_\mu = \sigma^2
\|\varphi_\mu\|^2_2$. In practice, the threshold values range
typically between three and four times the noise level $\sigma_\mu$
and in our experiments we will put $e_\mu = \sigma_\mu/2$. In short,
our constraints guarantee that the reconstruction will take into
account any pattern detected as significant by  any of the
$K$ transforms.
   
\subsubsection*{The Minimization Method}

We propose to solve (\ref{eq:l1-min}) using the method of hybrid
steepest descent (HSD) \citep{wave:yamada01}. HSD consists of building
the sequence
\begin{eqnarray}
 s^{n+1} = P(s^{n}) - \lambda_{n+1} \nabla_J(P(s^{n})); 
\end{eqnarray}
Here, $P$ is the $\ell_2$ projection operator onto the feasible set
$C$, $\nabla_J$ is the gradient of equation~\ref{eq:l1-min}, and
$(\lambda_{n})_{n \ge 1}$ is a sequence obeying $(\lambda_{n})_{n\ge
  1} \in [0,1] $ and $\lim_{ n \rightarrow + \infty } \lambda_{n} = 0$.

The combined filtering algorithm is:
\begin{enumerate}
\baselineskip=0.4truecm
\itemsep=0.1truecm
\item Initialize $L_{\max} = 1$, the number of iterations $N_i$, and
  $\delta_{\lambda} = \frac{L_{\max}}{N_i}$.
\item Estimate the noise standard deviation $\sigma$, and set $e_k =
  \frac{\sigma}{2}$.
\item For k = 1, .., $K$ calculate the transform: $\alpha^{(s)}_k
  = T_k s$.
\item Set $\lambda = L_{\max}$, $n = 0$, and $\tilde s^{n}$ to 0.
\item While $\lambda >= 0$ do
\begin{itemize}
\item $u = \tilde s^{n}$.
\item For k = 1, .., $K$ do
  \begin{itemize}
  \item Calculate the transform $\alpha_{k} = T_k u$.
  \item For all coefficients $\alpha_{k,l}$ do
     \begin{itemize}
     \item Calculate the residual $r_{k,l} = \alpha^{(s)}_{k,l} -
       \alpha_{k,l}$
       
     \item if $\alpha^{(s)}_{k,l}$ is significant and $ \mid r_{k,l}
       \mid > e_{k,l}$ then $\alpha_{k,l} = \alpha^{(s)}_{k,l}$
     \item $\alpha_{k,l} = sgn(\alpha_{k,l}) ( \mid \alpha_{k,l} \mid - \lambda)_{+}$.
     \end{itemize}
   \item $u = T_k^{-1} \alpha_{k}$
  \end{itemize}
\item Threshold negative values in $u$ and $\tilde s^{n+1} = u$.
\item $n = n + 1$, $\lambda = \lambda - \delta_{\lambda} $, and goto 5.
\end{itemize}
\end{enumerate}

\section*{Appendix 2: SMICA in the Spherical Wavelet Representation (wSMICA-S)}

A detailed derivation of SMICA and wSMICA can be found in~\citet{ica:Del2003} and \citet{starck:yassir05}.  
Details are given here showing how the Isotropic Undecimated Wavelet 
Transform on the Sphere described in Section~\ref{sect_wts} above can be used to extend SMICA to work 
in the wavelet domain on spherical data maps. We will refer to this extension as wSMICA-S.

\subsubsection*{wSMICA-S objective function}
In the linear mixture model~(\ref{model0}), $X$ stands for an array of observed spherical maps,   
$S$ is an array of spherical source maps to be recovered  and $N$ is an array of 
spherical noise maps. The mixing matrix $A$ achieves a pixelwise linear mixing of the 
source maps. The observations are assumed to have zero mean.  
Applying the above UWTS on both sides of~(\ref{model0}) does not affect 
the mixing matrix $A$ so that, assuming independent source and noise processes, 
the covariance matrix of the observations at scale $j$ is still  structured as  
\begin{equation}\label{structure}
R_w^X(j) = A R_w^S(j) A^{\dagger} +   R_w^N(j) 
\end{equation}
where  $R_w^S(j)$ and  $R_w^N(j)$ are the model diagonal spectral covariance
matrices in the wavelet representation of $S$ and $N$ respectively.
Provided estimates $\widehat{R}_w^X(j)$ of $R_w^X(j)$ can be obtained from the data,
our wavelet-based version of SMICA consists in minimizing the wSMICA-S
criterion:
\begin{equation}\label{Cost_wavelet}
 \Phi (\theta) =  \sum _{j=1}^{J+1}  \alpha_j \mathcal{D} \left( \widehat{R}_w^X(j), \,
    A R_w^S(j) A^{\dagger} + R_w^N(j) \right)
\end{equation}
 for some reasonable choice of the weights $\alpha_j$ and of the matrix mismatch measure $\mathcal{D}$, 
 with respect to the full set of parameters $\theta = (A,R_w^S(j), R_w^N(j) )$  or a subset thereof.  
 As discussed in~\citet{ica:Del2003} and \citet{starck:yassir05}, a good choice for $\mathcal{D}$ is
\begin{equation}
  \mathcal{D}_{KL} (R_1, R_2 ) 
  =
  \frac{1}{2} \Big( \mathrm{tr} (R_1R_2^{-1}) - \log\det (R_1R_2^{-1}) - m  \Big)
\end{equation}
which is the Kullback-Leibler divergence between two $m$-variate
zero-mean Gaussian distributions with covariance matrices $R_1$ and
$R_2$. With this mismatch measure, the wSMICA-S criterion is shown to be 
related to the likelihood of the data in a Gaussian model. 
We can resort to the EM algorithm to minimize~(\ref{Cost_wavelet}). 

In dealing with non stationary data or incomplete data, an 
attractive feature of  wavelet filters over the spherical harmonic transform is 
that they are well localized in the initial representation. Provided the wavelet filter  
response on scale $j$  is short enough compared to
data size and gap widths, most of the samples in the filtered
signal will then be unaffected by the presence of gaps.  
Using exclusively these samples yields an estimated covariance matrix $\widehat{R}_w^X(j)$ 
which is not biased by the missing data.  Writing the wavelet decomposition on $J$ scales of $X$ as 
\begin{equation}  
X(\vartheta, \varphi) = X_{J+1}^w (\vartheta, \varphi) + \sum_{j=1}^{J} X_{j}^w(\vartheta, \varphi)
\end{equation}
and denoting $l_j$ the size of  the set $\mathcal{M}_j$ of wavelet coefficients 
unaffected by the gaps at scale $j$,  the wavelet covariances  are simply estimated using
\begin{equation}
\widehat{R}_w^X (j) = \frac{1}{l_j }  \sum_{t \in  \mathcal{M}_j } X_j^w( \vartheta_t, \varphi_t )X_j^w( \vartheta_t, \varphi_t ) ^\dagger 
\end{equation}

The weights in the spectral mismatch~(\ref{Cost_wavelet}) should be
chosen to reflect the variability of the estimate of the corresponding
covariance matrix.  Since wSMICA-S uses wavelet filters with only limited overlap, 
in the case of complete data maps we follow the derivation in~\citet{ica:Del2003} 
and take $\alpha_j$ to be proportional to the number of spherical harmonic modes 
in the spectral domain covered at scale $j$.
In the case of data with gaps, we must further take into account that
only a fraction $\beta_j$ of the wavelet coefficients are unaffected so 
that  the  $\alpha_j$ should be modified in the same ratio. 

\subsubsection*{Source map estimation}

As a result of applying wSMICA-S, power densities in each scale are estimated  
for the sources and detector noise  along with the estimated mixing matrix.  These are
used in reconstructing the source maps \emph{via} Wiener filtering in
each scale: a coefficient $X_{j}^w(\vartheta, \varphi)$ 
is used to reconstruct the maps according to
\begin{equation}  \label{Wiener}
\widehat{S}_{j}^w(\vartheta, \varphi) = \big(\widehat{A} \adj \widehat{R}_w^N( j) ^{-1} \widehat{A} + \widehat{R}_w^S(j) ^{-1} \big)\inv \times \widehat{A} \adj \widehat{R}_w^N( j) ^{-1}  X_{j}^w(\vartheta, \varphi)
\end{equation}
In the limiting case where noise is small compared to signal
components, this filter reduces to 
\begin{equation}
 \widehat{S}_{j}^w(\vartheta, \varphi) 
  = (\widehat{A} \adj \widehat{R}_w^N( j) ^{-1} \widehat{A} )\inv \widehat{A} \adj \widehat{R}_w^N( j) ^{-1}  X_{j}^w(\vartheta, \varphi)  \label{Wiener1}
\end{equation}
Clearly, the above Wiener filter is optimal only for
stationary Gaussian processes.  For non Gaussian maps, such as given by the Sunyaev Zel'dovich  
effect,  better reconstruction can be expected from 
non linear methods.

\end{document}